\documentstyle[aps,prb,epsfig,floats]{revtex}

\begin{document}
\draft

\twocolumn[\hsize\textwidth\columnwidth\hsize\csname@twocolumnfalse\endcsname

\title{Diagrammatic self-energy approximations and the total particle number}
\author{Arno Schindlmayr,$^{1,}$\cite{email} P.\ Garc\'{\i}a-Gonz\'alez,$^2$
  and R.~W.\ Godby$^3$}
\address{$^1$Fritz-Haber-Institut der Max-Planck-Gesellschaft, Faradayweg
  4--6, 14195 Berlin-Dahlem, Germany}
\address{$^2$Departamento de F\'{\i}sica Fundamental, Universidad Nacional de
  Educaci\'on a Distancia, Apartado 60141,\\ 28080 Madrid, Spain}
\address{$^3$Department of Physics, University of York, Heslington, York YO10
  5DD, United Kingdom}
\date{\today}
\maketitle

\begin{abstract}
There is increasing interest in many-body perturbation theory as a practical
tool for the calculation of ground-state properties. As a consequence,
unambiguous sum rules such as the conservation of particle number under the
influence of the Coulomb interaction have acquired an importance that did not
exist for calculations of excited-state properties. In this paper we obtain a
rigorous, simple relation whose fulfilment guarantees particle-number
conservation in a given diagrammatic self-energy approximation. Hedin's
$G_0W_0$ approximation does not satisfy this relation and hence violates the
particle-number sum rule. Very precise calculations for the homogeneous
electron gas and a model inhomogeneous electron system allow the extent of the
nonconservation to be estimated.
\end{abstract}

]

\section{Introduction}

Many-body perturbation theory is a powerful method for studying interacting
electron systems, because the partial summation of self-energy diagrams allows
an efficient and systematically converging description of the dominant
scattering mechanisms.\cite{Fetter1971} In solid-state physics, Hedin's $GW$
approximation\cite{Hedin1965} includes dynamic screening in the random-phase
approximation and has been applied with great success to a large range of
materials.\cite{Aryasetiawan1998} While calculations have long focused on
electronic excitations, such as band structures,\cite{Hybertsen1985,Godby1986}
that are not normally accessible by variational mean-field schemes, there is
now increasing interest in using many-body perturbation theory also to obtain
ground-state properties like the charge density\cite{Rieger1998} or the total
energy\cite{Holm1999,Holm2000,Sanchez-Friera2000,Garcia-Gonzalez2000} in order
to circumvent well-known limitations of standard approximations in
density-functional theory.\cite{Hohenberg1964} Unlike the calculation of
excited states, which are given immediately by the pole structure of the
spectral function, this generally requires a multidimensional integration over
the hole part of the Green function, as in Galitskii and Migdal's expression
for the total energy.\cite{Galitskii1958} As a consequence, sum rules that
could hitherto be ignored have gained new prominence. The most important of
these is the conservation of particle number, i.e., the requirement that the
integral
\begin{equation}\label{Eq:number}
N = \frac{1}{2\pi i} \sum_{\sigma} \int\! d^3r \int\! d\omega\,
G_{\sigma\sigma }({\bf r},{\bf r};\omega) e^{i\omega\eta}
\end{equation}
over the diagonal elements of the Green function $G$ equals the true number of
electrons. Here $\sigma $ denotes the spin variable and $\eta $ is a positive
infinitesimal that forces the frequency contour to be closed across the upper
complex half plane.

In a seminal paper Baym and Kadanoff\cite{Baym1961} investigated the evolution
of nonequilibrium Green functions and derived a set of symmetry relations for
diagrammatic many-body approximations that guarantee the overall conservation
of particle number, total energy, and momentum under time-dependent external
perturbations. Baym\cite{Baym1962} later showed that a self-energy satisfying
all of these relations can be represented as the derivative $\Sigma =
\delta \Phi / \delta G$ of a generating functional $\Phi$, and that in this
case the Green function obtained self-consistently from Dyson's
equation\cite{Dyson1949} moreover yields the exact particle number. In
particular, this applies to the fully self-consistent $GW$ approximation, in
which both the Green function $G$ and the screened Coulomb interaction $W$ are
dressed by self-energy insertions in accordance with a self-consistent
solution of Dyson's equation.\cite{Baym1961} However, it has since become
clear that $\Phi$ derivability is a sufficient but not a necessary requirement
for the fulfilment of the particle-number sum rule. For instance, the
partially self-consistent $GW_0$ approximation, in which only the Green
function is updated self-consistently but the screened Coulomb interaction
remains undressed, is {\em not\/} $\Phi$ derivable but nevertheless produces
the exact particle number.\cite{Holm1997} On the other hand, without
self-consistency even in the Green function, the particle number is not, in
general, given correctly,\cite{Schindlmayr1997} but this computationally
efficient $G_0W_0$ approach still remains the preferred method for most
practical applications.

More complicated self-energy expressions like the cumulant
expansion\cite{Aryasetiawan1996} or the $T$ matrix\cite{Springer1998} have
already been successfully applied to solids, leading to an improved
description of satellite resonances. Like the $GW$ approximation, these
schemes are typically implemented without full self-consistency and are hence
not $\Phi$ derivable. In order to avoid expensive numerical tests in such
situations, it would be desirable to have clear diagrammatic criteria for the
fulfilment of the particle-number sum rule that could be checked {\em a
  priori\/} without actual calculations.

Unfortunately, Baym's proof, which is based on Luttinger's examination of the
exact theory\cite{Luttinger1960} and determines the volume of the Fermi sea
directly, cannot easily be extended, because it relies explicitly on the
existence of the generating functional $\Phi $. We therefore take a
different approach by describing the switching on of the Coulomb potential as
a time-dependent process that connects the noninteracting and the
corresponding interacting electron system on a finite time scale. In this way
we can examine the differential conservation laws and deduce a diagrammatic
symmetry relation for particle-number conservation {\em before\/} taking the
adiabatic limit. The theoretical framework is developed in Sec.\
\ref{Sec:conservation}. The non-self-consistent $G_0W_0$ approximation, which
violates this symmetry relation and does not conserve the particle number when
the interaction is switched on, deserves special attention owing to its
pre-eminent role in practical implementations. In Sec.\ \ref{Sec:numerical} we
therefore present very precise numerical calculations of the particle number
for the homogeneous electron gas and a model inhomogeneous system in order to
assess the quantitative deviation. Finally, in Sec.\ \ref{Sec:conclusions} we
summarize our conclusions. Atomic units are used throughout.

\section{Particle-number conservation}\label{Sec:conservation}

In order to connect the interacting electron system with the corresponding
noninteracting system, whose properties are supposedly known exactly, we
consider the Hamiltonian
\begin{equation}
\hat{H}(t) = \hat{H}_0 + e^{-\epsilon |t|} \hat{H}_1 \;.
\end{equation}
The one-body part $\hat{H}_0$ contains the kinetic energy as well as the
external potential $V_{{\rm ext}}$, and the Coulomb interaction $\hat{H}_1$ is
switched on exponentially with $\epsilon >0$. At large times, both in the past
and in the future, the Hamiltonian reduces to $\hat{H}_0$, which constitutes a
solvable problem. The noninteracting Green function $G_0$ is readily
constructed from the solutions of the single-particle Schr\"odinger equation
and yields the correct number of particles. On the other hand, at $t=0$ the
full Coulomb interaction is effective, and the Green function is defined
as\cite{Fetter1971}
\begin{equation}
G(x,x') = -i \frac{\langle \Psi | T[ \hat{\psi}(x) \hat{\psi}^{\dagger}(x') ]
  | \Psi \rangle}{\langle \Psi |\Psi \rangle} \;,
\end{equation}
where the shorthand notation $x\equiv ({\bf r},\sigma,t)$ indicates a set of
spatial, spin, and temporal coordinates, $| \Psi \rangle$ denotes the
ground-state wave function of the interacting electron system in the
Heisenberg picture, and $T$ is Wick's time-ordering operator that rearranges
the subsequent symbols in ascending order from right to left with a sign
change for every pair commutation. Furthermore, $\hat{\psi}^{\dagger}(x')$ and
$\hat{\psi}(x)$ represent the electron creation and annihilation operator in
the Heisenberg picture, respectively.

The unknown many-body wave function $| \Psi \rangle$ evolves from the
noninteracting ground state $| \Psi_0 \rangle$ and can formally be expressed
as $| \Psi \rangle = \hat{U}_\epsilon(0,-\infty) | \Psi_0 \rangle$,
where\cite{Fetter1971}
\begin{eqnarray}
\hat{U}_\epsilon(t,t') &=& \sum_{\nu=0}^\infty \frac{(-i)^\nu}{\nu!}
\int_{t'}^t \! dt_1 \cdots \int_{t'}^t \! dt_\nu \,
e^{-\epsilon (|t_1| + \cdots + |t_\nu|)}\nonumber\\
&&\times T[ \hat{H}_1(t_1) \cdots \hat{H}_1(t_\nu) ]
\end{eqnarray}
represents the time-development operator. With this definition the Green
function may be rewritten as
\begin{equation}\label{Eq:Green}
G(x,x') = -i \frac{\langle \Psi_0 | T[ \hat{S}_\epsilon \hat{\psi}(x)
  \hat{\psi}^\dagger(x') ] | \Psi_0 \rangle}{\langle \Psi_0 | \hat{S}_\epsilon
  | \Psi_0 \rangle}
\end{equation}
with $\hat{S}_\epsilon = \hat{U}_\epsilon(\infty,-\infty)$. At this stage the
Gell-Mann and Low theorem\cite{Gell-Mann1951} asserts that it is, in general,
permissible to take the adiabatic limit $\epsilon \to 0$. However, in the
following we continue to perform a time-dependent perturbation analysis for
finite $\epsilon$ and only take the adiabatic limit after establishing the
conservation criteria that apply during the transition. The time-ordered
products in Eq.\ (\ref{Eq:Green}) may be evaluated in the usual way by
invoking Wick's theorem,\cite{Wick1950} because the exponentials are scalar
functions and commute with the field operators. Hence the perturbative
treatment generates the standard series of connected and topologically
distinct Feynman diagrams,\cite{Feynman1949} made up of the noninteracting
Green function $G_0$ and the two-body Coulomb potential, but the latter now
acquires an additional prefactor and is given by 
\begin{equation}
v(x,x') = \frac{e^{-\epsilon |t|}}{|{\bf r}-{\bf r}'|} \delta(t-t') \;.
\end{equation}
The formal identity of the perturbation expansion in the time-dependent and
the adiabatic, time-independent case is a crucial result that forms the basis
of our discussion in this section.

For an analysis of the conservation properties we now follow Ref.\
\onlinecite{Baym1961} and write the perturbation series as
\begin{eqnarray}
\lefteqn{\int\! G_0^{-1}(x,x_1) G(x_1,x') \,dx_1}\nonumber\\
&=& \delta(x-x') - i \int\! v(x,x_1) G_2(x,x_1;x',x_1^+) \,dx_1
\;,\label{Eq:lefthand}
\end{eqnarray}
invoking the two-particle Green function $G_2$. The superscript $x^+$
indicates that a positive infinitesimal is added to the time variable to
ensure the proper ordering. An equivalent form is the adjoint equation of
motion
\begin{eqnarray}
\lefteqn{\int\! G(x,x_1) G_0^{-1}(x_1,x') \,dx_1}\nonumber\\
&=& \delta(x-x') - i \int\! G_2(x,x_1;x',x_1^+) v(x',x_1) \,dx_1
\;.\label{Eq:righthand}
\end{eqnarray}
The inverse noninteracting Green function is identical to the operator 
\begin{eqnarray}
G_0^{-1}(x,x') &=& \left( i \frac{\partial}{\partial t} + \frac{1}{2}
\nabla^2 - V_{\rm ext}({\bf r}) \right) \delta(x-x')\nonumber\\
&=& \left( -i \frac{\partial}{\partial t'} + \frac{1}{2}
{\nabla'}^2 - V_{\rm ext}({\bf r}') \right) \delta(x-x') \;,
\end{eqnarray}
so that after substracting Eq.\ (\ref{Eq:righthand}) from Eq.\
(\ref{Eq:lefthand}) we obtain
\begin{eqnarray}
\lefteqn{\left[ i \left( \frac{\partial}{\partial t} +
      \frac{\partial}{\partial t'} \right) + \frac{1}{2} \left( \nabla +
      \nabla' \right) \cdot \left( \nabla - \nabla' \right) \right]
      G(x,x')}\nonumber\\
&=& \left[ V_{\rm ext}({\bf r}) - V_{\rm ext}({\bf r}') \right]
      G(x,x')\label{Eq:difference}\\
&&- i \int\! \left[ v(x,x_1) - v(x',x_1) \right] G_2(x,x_1;x',x_1^+) \,dx_1
      \;.\nonumber
\end{eqnarray}
When we set $x'= x^+$, the terms on the right-hand side cancel, while the
left-hand side reduces to the differential conservation law for the particle
number 
\begin{equation}
\frac{\partial n({\bf r},t)}{\partial t} + \nabla \cdot {\bf j}({\bf r},t) = 0
\end{equation}
with the electron density $n({\bf r},t) = -i \sum_\sigma G(x,x^+)$ and current
${\bf j}({\bf r},t) = -\frac{1}{2} \sum_\sigma [ (\nabla - \nabla') G(x,x')
]_{x'=x^+}$. Thus whenever Eqs.\ (\ref{Eq:lefthand}) and (\ref{Eq:righthand})
are satisfied simultaneously, the total particle number is conserved while the
interaction is switched on. This does not depend on the value of $\epsilon$
and, in particular, remains true in the adiabatic limit, which can now be
taken, turning $G$ into the equilibrium Green function of the interacting
electron system.

By multiplying Eqs.\ (\ref{Eq:lefthand}) and (\ref{Eq:righthand}) with $G$
from the left and right, respectively, and then subtracting one from the
other, their mutual consistency can be stated in the more convenient form,
\begin{eqnarray}
\lefteqn{\int\! G(x,x_2) v(x_2,x_1) G_2(x_2,x_1;x',x_1^+) \,dx_1
  \,dx_2}\nonumber\\
&=& \int\! G_2(x,x_1;x_2,x_1^+) v(x_2,x_1) G(x_2,x') \,dx_1
  \,dx_2 \;,\label{Eq:consistency}
\end{eqnarray}
that may easily be verified by visual inspection of a given diagrammatic
approximation for the two-particle Green function. Evidently it is the same
criterion as derived by Baym and Kadanoff\cite{Baym1961} for particle-number
conservation under time-dependent external perturbations. This is no
coincidence, of course, because the termwise cancellation of diagrams on the
right-hand side of Eq.\ (\ref{Eq:difference}) is of purely topological origin
and does not depend on the mathematical properties of the constituent
propagators. Hence it is inconsequential whether, as in Refs.\
\onlinecite{Baym1961} and \onlinecite{Baym1962}, $G_0$ contains a
time-dependent perturbation while the interaction is constant or, as in the
physical situation considered here, the noninteracting Green function is
invariant under temporal translations while the Coulomb potential instead
acquires a time-dependent prefactor.

\begin{figure}[t!]
\epsfxsize=8cm \centerline{\epsfbox{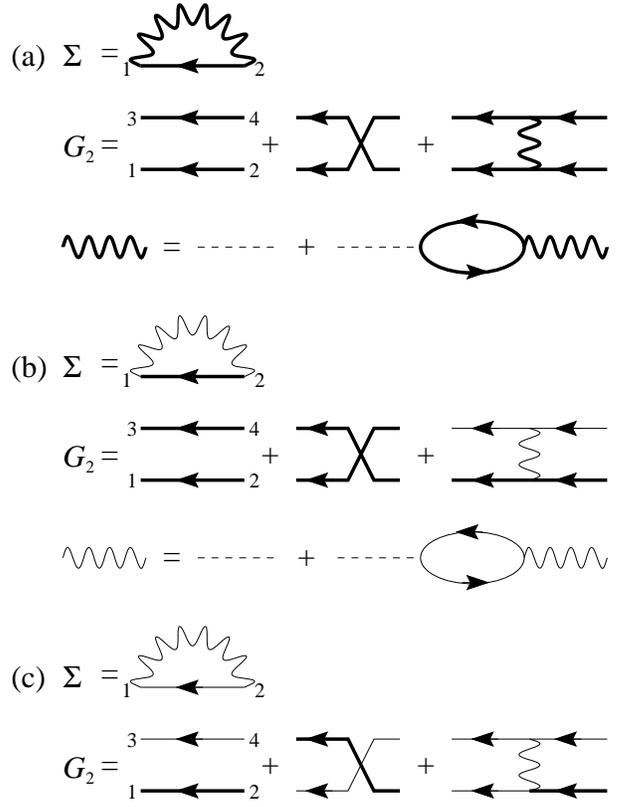}} \bigskip
\caption{Diagrammatic representation of the self-energy $\Sigma(x_1,x_2)$
and the corresponding two-particle Green function $G_2(x_1,x_3;x_2,x_4)$
in (a) the fully self-consistent $GW$ approximation, (b) the partially
self-consistent $GW_0$ approximation, and (c) the non-self-consistent $G_0W_0$
approximation.}
\label{Fig:diagrams}
\end{figure}

As an example we now consider the $GW$ approximation. The self-energy, when
applied with full self-consistency, is given by 
\begin{equation}
\Sigma(x,x') = i G(x,x') W(x^+,x') \;,
\end{equation}
where the screened Coulomb interaction $W$ takes the mathematical form of the
random-phase approximation but is evaluated using the dressed Green function
self-consistently derived from Dyson's equation. The diagrammatic
representation of $\Sigma$ is shown in Fig.\ \ref{Fig:diagrams}(a). The
corresponding two-particle Green function is obtained by comparing Dyson's
equation with the equation of motion (\ref{Eq:lefthand}), which yields the
identity
\begin{eqnarray}
\lefteqn{- i \int\! v(x,x_1) G_2(x,x_1;x',x_1^+) \,dx_1}\nonumber\\
&=& V_H(x) G(x,x') + \int\! \Sigma(x,x_1) G(x_1,x') \,dx_1 \;,
\end{eqnarray}
where $V_H(x) = -i \int\! v(x,x_1) G(x_1,x_1^+) \,dx_1$ indicates the Hartree
potential. The two-particle Green function corresponding to the $GW$
approximation for the self-energy is also displayed in Fig.\
\ref{Fig:diagrams}(a). It is easily seen that it satisfies the symmetry
relation (\ref{Eq:consistency}), which is essentially a horizontal left-right
symmetry for the building blocks of $G_2$, and hence conserves the total
particle number when the Coulomb interaction is switched on. Of course, this
result also follows from the existence of the generating functional
$\Phi$.\cite{Baym1962}

The partially self-consistent $GW_0$ approximation 
\begin{equation}
\Sigma(x,x') = i G(x,x') W_0(x^+,x') \;,
\end{equation}
in which the screened Coulomb interaction $W_0$ is evaluated with the
noninteracting Green function $G_0$, is {\em not\/} $\Phi$ derivable, which
would require an additional vertical mirror symmetry $G_2(x_1,x_3;x_2,x_4) =
G_2(x_3,x_1;x_4,x_2)$ in the diagrammatic structure of the two-particle Green
function that has been lost in the transition from full to partial
self-consistency. Nevertheless, $G_2$, shown in Fig.\ \ref{Fig:diagrams}(b),
still obeys the consistency relation (\ref{Eq:consistency}) and hence
guarantees the correct total particle number, as previously confirmed by
explicit integration of the spectral function.\cite{Holm1997} In contrast, the
non-self-consistent $G_0W_0$ approximation
\begin{equation}\label{Eq:G0W0}
\Sigma(x,x') = i G_0(x,x') W_0(x^+,x')
\end{equation}
leads to a two-particle Green function with lower internal symmetry, displayed
in Fig.\ \ref{Fig:diagrams}(c), that no longer satisfies Eq.\
(\ref{Eq:consistency}), implying an incorrect total particle number. The
quantitative deviation is investigated in the following section. In a similar
manner, the conservation properties of other diagrammatic self-energy
approximations are easily established by an inspection of the underlying
two-particle Green function.

\section{Numerical results}\label{Sec:numerical}

In the previous section we proved that the $GW$ and $GW_0$ approximations
conserve the particle number for an arbitrary electron system when the Coulomb
interaction is switched on, in contrast to $G_0W_0$. This, coupled with their
superior performance in ground-state total-energy
calculations,\cite{Holm1999,Garcia-Gonzalez2000} might be thought to suggest
that the $G_0W_0$ approach is useless if one is interested in ground-state
properties. However, a many-body calculation at only the $G_0W_0$ level is
already sufficient to correct typical limitations of mean-field
density-functional theories, such as their inaccuracy in highly inhomogeneous
systems or their failure to describe van der Waals
forces.\cite{Garcia-Gonzalez2001} Moreover, the Green function arising from a
$G_0W_0$ calculation may be used as input in the variational Luttinger and
Ward functional,\cite{LuttingerWard1960} and prospective calculations suggest
that this is an excellent approach for calculating total
energies.\cite{Almbladh1999} Since these methods are more amenable to
applications in complex systems than the fully or partially self-consistent
$GW$ approximations, it is important to determine whether the underlying
violation of the particle-number sum rule in the $G_0W_0$ framework is small
enough to be safely ignored.

There are some indications that such an error is indeed fairly small for the
homogeneous electron gas at metallic densities,\cite{Holm1997} a Hubbard model
system,\cite{Schindlmayr1997} and typical semiconductors.\cite{Rieger1998}
Here, bearing in mind that many-body total-energy calculations are intended to
be used in extreme situations where standard implementations of
density-functional theory fail, we present numerical results for thin jellium
slabs, whose most relevant feature is the strong inhomogeneity of the
electron-density profile, as well as for the homogeneous electron gas over a
wide range of densities.

Our concern is the evaluation of the particle-number difference,
\begin{equation}\label{Eq:varnum}
\delta N = \frac{-i}{\pi} \int_{-\infty}^{+\infty} \! d\omega\, \,\mbox{tr}
\left[ G(\omega) - G_0(\omega) \right] \;,
\end{equation}
where tr denotes the spatial trace (we omit the explicit spatial variables for
clarity and also consider only spin-unpolarized systems). Since both $G$ and
$G_0$ behave as $1 / \omega$ for large frequencies, we can apply Cauchy's
theorem and write Eq.\ (\ref{Eq:varnum}) alternatively as
\begin{equation}\label{Eq:varnumbis}
\delta N = \frac{1}{\pi} \int_{-\infty}^{+\infty} \! d\omega\, \,\mbox{tr}
\left[ G(\mu + i \omega) - G_0(\mu_0 + i \omega) \right] \;,
\end{equation}
where $\mu$ and $\mu_0$ are the chemical potentials of the interacting and the
noninteracting system, respectively, which correspond, by definition, to the
position of the pole of the Green function at the Fermi surface. As the
characteristic sharp structure of $G(\omega)$ (quasiparticle peaks and
satellites) does not appear in the analytic continuation $G(\mu+i\omega)$,
Eq.\ (\ref{Eq:varnumbis}) is preferred for numerical integration. We hence
follow some of the ideas suggested by Rojas {\em et al.}\cite{Rojas1995} and
work exclusively in an imaginary time and frequency representation. An
accurate evaluation of Eq.\ (\ref{Eq:varnumbis}) furthermore requires a
treatment of the high-frequency tails of $G$, which can be done easily with
the numerical procedures described in Ref. \onlinecite{Steinbeck2000}.

For the homogeneous electron gas, an analytic expression exists for the
noninteracting Green function $G_0(r,i\tau)$ in real space and imaginary
time,\cite{Schindlmayr2000} while the screened Coulomb interaction
$W_0(k,i\omega)$ in the random-phase approximation is given analytically in
reciprocal space by the dynamic Lindhard function.\cite{Lindhard1954} The
evaluation of the self-energy according to $\Sigma(r,i\tau) = i G_0(r,i\tau)
W_0(r,i\tau)$ therefore only requires the numerical Fourier transform
$W_0(k,i\omega) \to W_0(r,i\tau)$. It is this largely analytic approach that
makes the present calculation especially precise.

At this stage we remark that the self-energy given by Eq.\ (\ref{Eq:G0W0}) has
the same analytic structure as the underlying Green function $G_0$, i.e., the
poles of $\Sigma(\omega)$ are located in the upper (lower) complex half-plane
for energies smaller (larger) than $\mu_0 = \frac{1}{2} k_{\rm F}^2$, where
$k_{\rm F}$ denotes the Fermi wave vector. As a consequence, an inconsistency
arises because the true self-energy should have a polar structure identical to
the {\em interacting\/} Green function with the chemical potential $\mu$. The
self-energy must therefore be appropriately shifted along the real frequency
axis. In the imaginary time/frequency representation, this shift is
automatically included in the backward transform
\begin{equation}
\Sigma(\mu + i \omega) = -i \int_{-\infty }^{+\infty } \! d\tau\,
\Sigma(i\tau) e^{-i\omega\tau} \;.
\end{equation}
The calculation of $\Sigma(\mu+i\omega)$, therefore, does not require an
advance knowledge of $\mu$, which can now be obtained from the relation $\mu =
\mu_0 + \Sigma(k_{\rm F},\mu)$.

\begin{figure}[t!]
\epsfxsize=7.8cm \centerline{\epsfbox{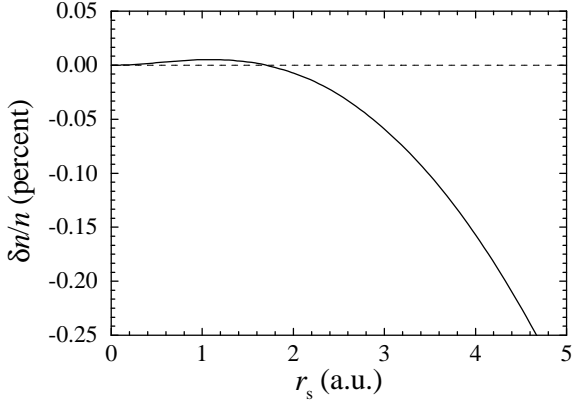}} \bigskip
\caption{Violation of the particle-number sum rule for the homogeneous
  electron gas in the $G_0W_0$ approximation. The relative error in the
  density is always negative and of the order of 0.1\% in the range of
  metallic densities.}
\label{Fig:errhom}
\end{figure}

Finally, the interacting Green function is calculated in reciprocal space
according to
\begin{equation}
G(k,\mu+i\omega) = \frac{1}{i \omega - \frac{1}{2} k^2 - \Sigma(k,\mu+i\omega)
  + \mu} \;.
\end{equation}
In the same representation, the noninteracting Green function is given by
\begin{equation}
G_0(k,\mu_0+i\omega) = \frac{1}{i \omega - \frac{1}{2} k^2 + \mu_0} \;,
\end{equation}
so that the density variation is readily obtained from
\begin{equation}
\delta n = \int\! \frac{d^3k}{(2\pi)^3} \int_{-\infty}^{+\infty} \!
\frac{d\omega}{\pi} \left[ G(k,\mu+i\omega) - G_0(k,\mu_0+i\omega) \right] \;.
\end{equation}
In Fig.\ \ref{Fig:errhom} the relative devation $\delta n / n_0$ from the
exact density is displayed as a function of the Wigner-Seitz radius $r_{\rm
  s}$. In the high-density region $r_{\rm s} < 1.8$ the particle number is
slightly overestimated ($<0.01\%$), while it is underestimated for lower
densities. In the range of metallic densities this underestimation is of the
order of 0.1\%, but the error becomes increasingly important in the dilute
limit ($-1.7\%$ for $r_{\rm s} = 10$ and $-6.1\%$ for $r_{\rm s} = 20$).

As pointed out above, it is also of interest to investigate the error
resulting from the $G_0W_0$ method in the total number of particles for a
strongly inhomogeneous system. The model we have chosen is a thin jellium slab
with a background density $n_0 = ( \frac{4}{3} \pi r_{\rm s}^3)^{-1}$ and
width $L$. The slab is bounded by two infinite planar walls, so that, if
charge neutrality is assumed, the system is fully characterized by the lengths
$r_{\rm s}$ and $L$. In this case, $G_0$ corresponds to the Kohn-Sham system
obtained self-consistently with the local-density approximation (LDA) for the
exchange-correlation potential $V_{\rm xc}$, as is typically done in practical
{\em ab initio\/} calculations.

With $z$ chosen as the coordinate perpendicular to the planar walls, the
translational symmetry of the system in the $xy$ plane allows an efficient
semianalytic evaluation of the relevant propagators. The screened Coulomb
interaction is given by $W_0 = \epsilon_0^{-1} v$, where $\epsilon_0$ denotes
the dielectric function in the random-phase approximation. The latter is
calculated as $\epsilon_0(k,i\omega)_{\alpha\beta}$ in the basis
$\zeta_\alpha(z) \exp(i{\bf k} \cdot \mbox{\boldmath $\rho$}) / \sqrt{S}$.
Here $\zeta_\alpha(z)$ is a set of cosine functions, ${\bf k} = (k_x,k_y)$ and
$\mbox{\boldmath $\rho$} = (x,y)$ denote the two-dimensional momentum and the
position vector in the $xy$ plane, respectively, and $S$ is the slab
surface. The matrix elements can be calculated analytically in terms of the
scalar products $\langle \zeta_\alpha \phi_n | \phi_m
\rangle$,\cite{Garcia-Gonzalez2001,Eguiluz1985} where $\phi_n(z) \exp(i{\bf k}
\cdot \mbox{\boldmath $\rho$}) / \sqrt{S}$ are the single-particle eigenstates
of the Kohn-Sham Hamiltonian $h_{\rm KS}$. The matrix elements
$v(k)_{\alpha\beta}$ of the Coulomb potential are likewise obtained
analytically. The screened Coulomb interaction is then easily calculated by a
matrix inversion for each value of $k$, and the real-space representation is
given by expanding
\begin{eqnarray}
W_0(\rho,z,z';i\tau)
&=& i\sum_{\alpha,\beta} \int\! \frac{d^2k}{(2\pi)^2} \int_{-\infty}^{+\infty}
\frac{d\omega}{2\pi} e^{i(\omega\tau + {\bf k} \cdot \mbox{\scriptsize
    \boldmath $\rho$})}\nonumber\\
&&\times \zeta _\alpha(z) \zeta_\beta(z') W_0(k,i\omega)_{\alpha\beta} \;.
\end{eqnarray}
The Green function $G_0(\rho,z,z';i\tau)$ is readily calculated from the
Kohn-Sham eigenstates, and by employing Eq.\ (\ref{Eq:G0W0}) we obtain the
self-energy in real space and imaginary time as well as, eventually, its
representation $\Sigma(k,\mu+i\omega)_{nm}$ in the Kohn-Sham basis set. The
presence of infinite confining walls implies a quick convergence with respect
to the number of cosine and Kohn-Sham wave functions used in the
calculation. The convergence is further accelerated by the analytic treatment
of the asymptotic time and frequency tails of all operators.

The Green function is calculated in the basis of Kohn-Sham eigenstates
according to
\begin{eqnarray}
G(k,\mu+i\omega) &=& \left[ i\omega - h_{\rm KS}(k) -
 \Sigma(k,\mu+i\omega) \right.\nonumber\\
&&+ \left. V_{\rm xc}(k) + \mu \right]^{-1}
\end{eqnarray}
by a matrix inversion in the indices $nm$. Finally, the variation of the
number of particles per surface unit is given by
\begin{eqnarray}
\frac{\delta N}{S} &=& \sum_m \int\! \frac{d^2k}{(2\pi)^2}
\int_{-\infty}^{+\infty} \frac{d\omega}{\pi} \left[
  G(k,\mu+i\omega)_{mm} \right.\nonumber\\
&&- \left. G_0(k,\mu_0+i\omega)_{mm} \right] \;,
\end{eqnarray}
where we have used the invariance of the trace with respect to any
wave-function representation.

\begin{figure}[t!]
\epsfxsize=7.8cm \centerline{\epsfbox{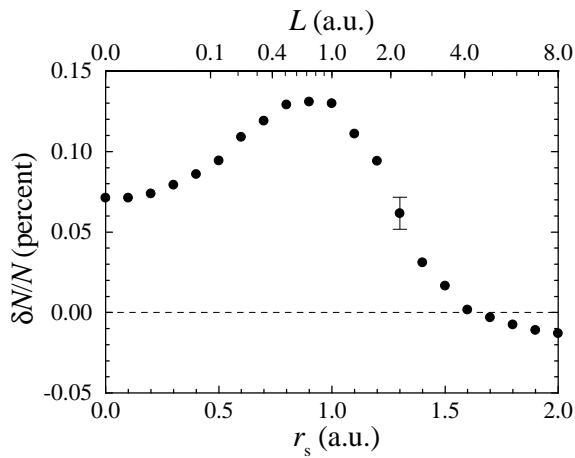}} \bigskip
\caption{Relative violation of particle number in the $G_0W_0$ approximation
  for thin jellium slabs of fixed 2D density $n^{\rm 2D} = 3/4\pi$ as a
  function of their thickness $L$ (and the corresponding 3D density parameter
  $r_{\rm s}$). A typical error bar is reported.}
\label{Fig:errmod}
\end{figure}

In Fig.\ \ref{Fig:errmod} we plot the relative deviation of the particle
number $\delta N / N$ in the $G_0W_0$ approximation for several configurations
of the model system, keeping the exact number of particles per surface unit
$n^{\rm 2D} = n_0 L = L / (\frac{4}{3} \pi r_{\rm s}^3)$ constant. The limit
$L \to 0$ thus corresponds to a two-dimensional (2D) homogeneous electron gas
with density $n^{\rm 2D}$. Over the wide variation of the degree of
homogeneity shown in the figure, it is seen that $\delta N / N$ remains of
similar magnitude as in the homogeneous case ($\lesssim 0.2\%$). This
observation remains true for other 2D densities inside the range [0.1,1].

\section{Conclusions}\label{Sec:conclusions}

In this paper we have rigorously obtained a general criterion which allows,
by simple inspection, to verify whether a diagrammatic self-energy
approximation satisfies the particle-number sum rule for an interacting
electron system. As an application, we have demonstrated that the so-called
$G_0W_0$ method does not yield the correct particle number, generalizing the
conclusions of a previous analytic study for a Hubbard model Hamiltonian
defined only on a discrete lattice.\cite{Schindlmayr1997} Thus
this limitation of the $G_0W_0$ approximation has been fully confirmed
for arbitrary electron systems. By performing a very precise integration of
the spectral function, we have furthermore calculated the size of the error in
the $G_0W_0$ particle number in two simple, but very distinct, families of
electron systems. The error becomes large only outside the range of densities
of physical interest.

\acknowledgements

The authors thank Professor C.-O.\ Almbladh and Dr.\ J.~E.\ Alvarellos for a
thorough reading of the manuscript and for valuable discussions. This work was
funded in part by the EU through the NANOPHASE Research Training Network
(Contract No.\ HPRN-CT-2000-00167), the Spanish Education Ministry DGESIC
Grant No.\ PB97-1223-C02-02, and by the Deut\-scher Aka\-de\-mi\-scher
Aus\-tausch\-dienst and the British Council under the British-German Academic
Research Collaboration (ARC) program.

\end{document}